\documentstyle[aps,preprint,epsf]{revtex}

\begin{document}

\renewcommand{\pl}{\partial}
\renewcommand{\arg}{(\vec x, \,\vec p, \,t)}

\draft
\preprint{TPR-95-19}

\title{
   Subthreshold Antiproton Spectra \\
   in Relativistic Heavy Ion Collisions
}

\author{Richard Wittmann and Ulrich Heinz}

\address{
   Institut f\"ur Theoretische Physik, Universit\"at Regensburg,\\
   D-93040 Regensburg, Germany
}
\date{\today}

\maketitle

\begin{abstract}
  We study the structure of antiproton spectra at extreme subthreshold
  bombarding energies using a thermodynamic picture. Antiproton
  production processes and final state interactions are discussed in
  detail in order to find out what can be learned about these processes
  from the observed spectra.
\end{abstract}


\section{Introduction}\label{sec1}

There exist numerous examples for the production of particles in heavy
ion collisions at bombarding energies well below the single
nucleon-nucleon threshold \cite{Mo}. This phenomenon indicates
collective interactions among the many participating nucleons and thus
is expected to give information about the hot and dense matter formed
in these collisions.  At beam energies around 1~GeV per nucleon the
most extreme of these subthreshold particles is the antiproton.
Therefore, it is believed to be a very sensitive probe to collective
behaviour in nucleus-nucleus collisions.

However, presently neither the production mechanism nor the final
state interactions of antiprotons in dense nucleonic matter are well
understood. The antiproton yields measured at GSI and BEVALAC
\cite{Schr,Sh} seem to be described equally well by various
microscopic models using different assumptions about the production
mechanism and particle properties in dense nuclear matter
\cite{TCMM,LK,Bat,Sp}. This ambiguity raises the question which kind of
information can be really deduced from subthreshold $\bar p$ spectra.
In this paper we use a simple thermodynamic framework as a background
on wich we can systematically study the influence of different
assumptions on the final $\bar p$ spectrum.

In the next section we will focus on the production process. Following
a discussion of the final state interactions of the antiproton in dense
hadronic matter in Section~\ref{sec3}, we use in Section~\ref{sec4} a
one-dimensional hydrodynamic model for the exploding fireball to
clarify which features of the production and reabsorption mechanisms
should survive in the final spectra in a dynamical environment.
We summarize our results in Section \ref{sec5}.

\section{Production of Antiprotons in Heavy Ion Collisions}
\label{sec2}

\subsection{The Antiproton Production Rate}
\label{sec2A}

Unfortunately very little is known about the production mechanism for
antiprotons in dense nuclear matter. Therefore, we are forced to use
intuitive arguments to obtain a plausible expression for the
production rate. As commonly done in microscopic models \cite{BG} we
consider only two body collisions and take the experimentally measured
cross sections for $\bar p$ production in free NN collisions as input.
The problem can then be split into two parts: the distribution of the
two colliding nucleons in momentum space and the elementary cross
sections for antiproton production. The procedure is later generalized
to collisions among other types of particles (Section~\ref{sec2C})
using phase space arguments. Formally, the antiproton production rate
$P$, i.e. the number of antiprotons produced in the space-time cell
$d^4 x$ and momentum space element $d^3 p$, is given by \cite{KMR}
 \begin{equation}
   P = \sum_{i,j} \int \, ds \, 2 \, w(s) \,
   \frac{d^3 \sigma_{ij\rightarrow \bar p}}{d^3 p}(s) \,  \int d\omega_i\,
   d\omega_j \, \delta(s-(p_i+p_j)^2) \,  f_i  \, f_j \,.
 \label{P}
 \end{equation}
In this expression the summation includes all incoming particle species $i$
and $j$ with 4-momenta $p_i$ and $p_j$ and phase space distributions $f_i$ and
$f_j$, respectively. The integration is over the invariant collision energy $s
= (p_i + p_j)^2$ and the phase space of the two colliding particles:
 \begin{equation}
    d\omega_i \equiv \frac{d^4 p_i}{(2\pi)^3} \,
                     \delta(p_i^2 - m_i^2) \, \Theta(p_i^0) \,.
 \label{domega}
 \end{equation}
Finally, $w(s, m_i, m_j) = \sqrt{ \left( s - ( m_i -m_j)^2\right) \,
\left(s - (m_i + m_j)^2 \right)} $ is the well known flux factor, and
${d^3 \sigma_{ij\rightarrow \bar p}}(s)/{d^3 p}$ the differential
$\bar p$ production cross section. The influence of the shape of the
phase space distributions and of the differential production cross
sections will now be investigated in more detail, beginning with the
case of $\bar p$ production by $NN$ collisions.

\subsection{Phase Space Distributions and Parametrisation of the
            Differential Production Cross Sections}
\label{sec2B}

Assuming that the momentum distribution of the nucleons gradually
evolves from $\delta$-peaks at the momenta of the colliding nuclei to
a relativistic Maxwell-Boltzmann distribution, an analytical
expression for the phase space distribution $f_{i,j}$ was derived by
Sch\"urmann {\em et al.} in Ref.~\cite{SHP}. We use this model as the
basis for a numerical evaluation of the probability $\lambda(s,t)$ of
finding two nucleons at a center-of-mass energy $\sqrt s$ at time $t$
(the second integral on the right hand side of Eq.~(\ref{P})).
Fig.~\ref{Bild1} shows $\lambda(s,t)$ at $t=0.2,\,2,\,5\,$~fm/c after
the nuclei have started overlapping, assuming a bombarding energy of
1~GeV/A. Although the initial $\delta$-like distribution at $s=5.5
\mbox{ GeV}^2$ widens quite rapidly, the extreme states needed for
antiproton production ($ s \geq (4 m)^2 = 14.1 \mbox{GeV}^2$) are
populated substantially only after about 5~fm/c. At this moment the
distribution is already nearly a relativistic Maxwell-Boltzmann
distribution (dashed line) which maximizes the $\bar p$ production rate
in our approach. This fact motivates us to neglect pre-equilibrium
production and to switch on $\bar p$-production only when the system
has reached local thermal equilibrium. Similiar conclusions have been
drawn from the results of a microscopic kinetic model for the phase
space evolution \cite{TCMM}.  There it was demonstrated that only
nucleons which have suffered at least several collisions are effective
in $\bar p$ production.

%
%

There are no experimental data for the nucleonic processes
 \[  p + p \rightarrow p + p + p + \bar p \]
 \[  n + p \rightarrow n + p + p + \bar p \]
close to the production threshold. As already pointed out in
\cite{Gie}, different extrapolations of the data at higher energies
down to threshold change the $\bar p$-yield by orders of magnitude.
However, we will show that not only the total yield, but also the
shape of the spectrum is very sensitive to the threshold behaviour. We
choose for the production cross section the parametrization
 \begin{equation}
   \sigma(s) = \sigma_0 \,
               {\left(\sqrt{s} - \mbox{4}\,m\right)^{\alpha}
                \over w^2(s)} \, ,
 \label{sigtot}
 \end{equation}
where the parameter $\alpha$ models the shape of the cross section near
threshold. The normalization $\sigma_0$ is determined by the experimentally
measured point closest to threshold. If the behaviour of $\sigma(s)$ were
dominated by the available phase space for the outgoing particles (this is the
fundamental assumption in Fermi's statistical model \cite{Mi}), then $\alpha =
7/2$.

In order to obtain from the total cross section (\ref{sigtot}) a
formula for the differential cross section, we assume like others that
the momentum distribution of the produced particles is mainly governed
by phase space. This leads to the simple relationship
 \begin{equation}
  \frac{d^3 \sigma_{ij\rightarrow NNN\bar p}} {d^3 p} (s) \sim
        \frac{ R_3 (P - p_{\bar p})} {R_4(P)}
        \ \sigma_{ij\rightarrow NNN\bar p}(s) \,.
 \label{sigdiff}
 \end{equation}
Here $R_n$ is the volume the $n$-particle phase space, which can be
given analytically in the non-relativistic limit \cite{Mi}. $P$ is the
total 4-momentum of the 4-particle final state, which reduces to
$(\sqrt{s}, \,0)$ in the CMS of the four particles.

\begin{figure}
  \epsfysize8cm
  \epsfbox{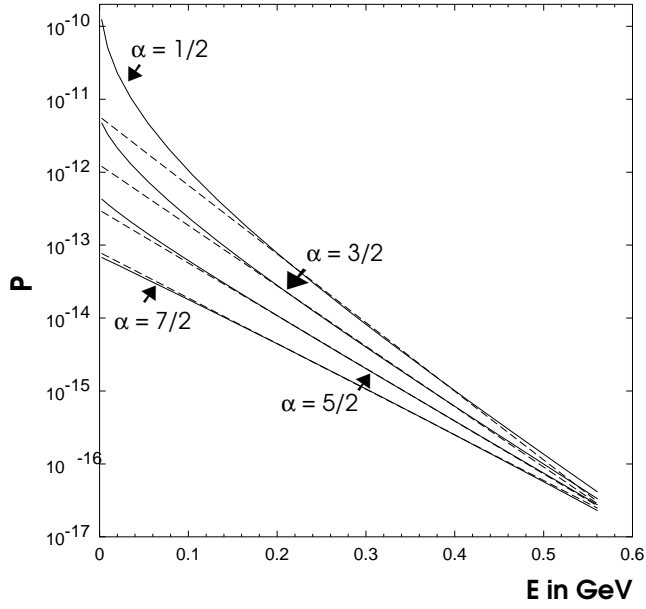}
  \vskip 0.4cm
  \caption{
    Antiproton production spectrum for different threshold behaviour
    of the elementary production process ($x = \frac{1}{2},
    \frac{3}{2}, \frac{5}{2}, \frac{7}{2}$ from top to bottom). }
  \label{Spektrum}
\end{figure}

Inserting (\ref{sigtot}) and (\ref{sigdiff}) into (\ref{P}) and using
thermal phase space distributions, the production spectrum can be
evaluated numerically. As seen in Fig.~\ref{Spektrum}, the sensitivity
on the threshold power $\alpha$ is particularly pronounced for
low-energy antiprotons. The strong deviation at low $\bar p$ energies
from a thermal shape (dashed lines) for small values of $\alpha$ is
most remarkable.

%
%

Using the Fermi value $\alpha = 7/2$ for the cross section, an
analytical expression for the production rate can be obtained:
  \[ P \sim \rho_i\,\rho_j\, e^{-\sqrt{s_{\rm min}(p)} / T} . \]
Here $\rho_{i,j}$ are the densities of the incoming particles,
$\sqrt{s_{\rm min}(p)}$ is the threshold for producing an antiproton
with momentum $p$, and $T$ is the local temperature of the
environment. Clearly the rate $P$ is proportional to the product of
the densities of the two colliding particle species $i$ and $j$. More
importantly, this formula shows the extreme sensitivity on the
production threshold and on the medium temperature which both appear
in the exponent of the exponential function.

\subsection{Production by Resonances and Mesons}
\label{sec2C}

The production of resonances as intermediate energy storage has been
pointed out as an important mechanism for $\bar p$ production in
\cite{KD}. Because no experimental data are available for antiproton
production by baryonic resonances, we follow common strategy \cite{KD}
and assume that the cross section $\sigma_{ij\rightarrow NNN\bar
p}(s)$ is independent of the internal state of excitation of the
colliding baryons in our thermal picture. The consequences of this
assumption are quite obvious. While the distance to the
$\bar p$-threshold is reduced by the larger rest mass of the
resonances, the mean velocity of a heavy resonance state in a thermal
system is smaller than that of a nucleon. Both factors counteract each
other, and indeed we found that the total rate $P$ is not strongly
changed by the inclusion of resonances.

The role of pionic intermediate states for $\bar p$-production in
$pp$-collisions was pointed out by Feldman \cite{Feld}. As mesons are created
numerously in the course of a heavy ion collision, mesonic states gain even
more importance in this case. In fact, Ko and Ge \cite{KG} claimed that
$\rho\rho \rightarrow p \bar p$ should be the dominant production channel.
Relating the $\rho\rho$-production channel to the $p\bar p$ annihilation
channel \cite{KG} by
 \begin{equation}
    \sigma_{\rho\rho \rightarrow p\bar p}(s) =
    \left( \frac{2}{2 S + 1} \right)^2 \
    \left( \frac{s - 4 m}{s - 4 m_{\rho}} m\right) \
    \sigma_{p\bar p \rightarrow \rho\rho}(s)
 \label{sigmarhorho}
 \end{equation}
where $S=1$ is the spin factor for the $\rho$, the production rate can be
calculated straightforwardly from Eq.~(\ref{P}):
 \begin{equation}
    P  =  {g_{\rho}^2 \over 6} \,
          \frac{\pi T}{(2\pi)^5} \,
          \frac{16}{9\pi}\,  w(4 E^2, m, m) \
          \sigma_{p\bar p \rightarrow\rho\rho}(4 E^2) \,
          K_1\left(\frac{2E}{T}\right)\,.
 \label{Prhorho}
 \end{equation}
The spin-isospin factor of the $\rho$ is $g_\rho = 9$, and $E$ is the
energy of the produced antiproton. The modified Bessel function $K_1$
results from the assumption of local thermal equilibration for the
$\rho$-distribution. Expanding the Bessel function for large values of
$2E/T$ we see that the "temperature" $T_{\bar p}$ of the
$\bar p$-spectrum is only half the medium temperature: $T_{\bar p} =
\frac{1}{2} T$.

Finally, the meson-baryon collision process
 \begin{equation}
   m_i+B_j \rightarrow B_k+p+\bar p
 \end{equation}
remains as the last prototype of collisions with two incoming
particles. We choose a parametrisation for the cross section which
reflects its qualitative behaviour \cite{KD}:
 \begin{equation}
   \sigma(ij\rightarrow N N \bar p)(s) = \sigma^0_{ij} \, s_0 \,
        \frac{ s- s_0} {w^2(s,m_i, m_j)}
 \label{sigPi}
 \end{equation}
Comparing this form with measured data on $\pi^- p \rightarrow n p
\bar p$ collisions \cite{LB}, a value of $\sigma^0_{ij} = 0.35$~mb is
obtained. Due to the threshold behaviour of Eq.~(\ref{sigPi}) and the
rather large value of $\sigma^0_{ij}$, it turns out \cite{Diss} that
this process is by far the most important one in a {\em chemically}
equilibrated system. However, this chemical equilibration -- if
achieved at all -- is reached only in the final stages of the heavy
ion collision when cooling has already started. So it is by no means
clear whether the dominance of the meson-baryon channel remains valid
in a realistic collision scenario. This point will be further
discussed in Section~\ref{sec4}.

\section{Final State Interaction of the Antiproton}
\label{sec3}

Once an antiproton is created in the hot and dense hadronic medium, its state
will be modified by interactions with the surrounding particles. Two
fundamentally different cases have to be distinguished: elastic scattering,
which leads to a reconfiguration in phase space, driving the momentum
distribution towards a thermal one with the temperature of the
surrounding medium, and annihilation. Each process will be considered
in turn.

\subsection{Elastic Scattering}
\label{sec3A}

The time evolution of the distribution function $f(p,t)$ is generally
described by the equation \cite{Ka}
 \begin{equation}
   f(p,t_2) = \int w(p, p'; t_2, t_1) \, f(p', t_1) \, dp'
 \label{f1}
 \end{equation}
where $w(p, p'; t_2, t_1)$ is the transition probability from momentum state
$p'$ at time $t_1$ to state $p$ at $t_2$. Because the number density of
antiprotons is negligible compared to the total particle density in the
system, the evolution of $f(p,t)$ can be viewed as a Markov process. Assuming
furthermore that the duration of a single scattering process $\tau$ and the
mean free path $\lambda$ are small compared to the typical time scale $\delta
t$ and length scale $\delta r$ which measure the variation of the
thermodynamic properties of the system,
  $$
    \tau \ll \delta t, \quad \lambda \ll \delta r\, ,
  $$
Eq.~(\ref{f1}) can be transformed into a master equation. Considering
the structure of the differential $p\bar p$ cross section one notices
that in the interesting energy range it is strongly peaked in the
forward direction \cite{Co,Br,Ei}. Therefore, the master equation can
be approximated by a Fokker-Planck equation \cite{Ch}:
 \begin{equation}
   \frac{\pl f(p,t)}{\pl t}= - \frac{\pl}{\pl p} \, A(p) f(p,t) +
   \frac{1}{2} \frac{\pl^2}{\pl^2 p} \,D(p) f(p,t) \,.
 \label{FPG}
 \end{equation}
For the evaluation of the friction coefficient $A$ and the diffusion
coefficient $D$ we follow the treatment described by Svetitsky
\cite{Sv}. For the differential cross section we took a form suggested
in \cite{Co}:
 \begin{equation}
   \frac{d\sigma}{dt} = (b + c \lambda) \,
   \frac{J_1^2\left[(a+\lambda)\sqrt{|t|} \right]} {|t|}\,,
 \end{equation}
where the constants $a,\,b,\,c$ are fitted to experimental data.
$\lambda$ is the wave number of the relative motion, $t$ the
transferred four momentum and $J_1$ a spherical Bessel function. The
essential result \cite{Diss} is that the coefficients $A$ and $B$ are
nearly independent of the $\bar p$ momentum and satisfactorily fulfill
the Einstein relation
 \begin{equation}
    D = m\, T \, A\,.
 \label{Einstein}
 \end{equation}

\begin{figure}
 \epsfxsize=10cm
 \epsfbox{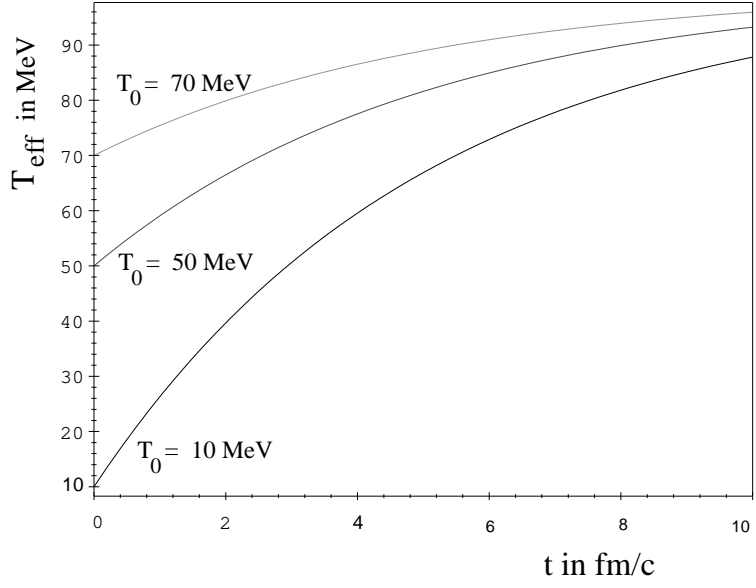}
 \vskip 0.4cm
 \caption{
   Effective temperature $T_{\rm eff}$ for three Maxwell distributions
   with initial temperatures $T_0 =$~10~MeV, 50~MeV and 70~MeV,
   respectively.}
 \label{Abb33}
\end{figure}

The effect of these coefficients can be visualized by examining the
time evolution of a non-relativistic Maxwell-Boltzmann distribution
whose initial slope $T_0$ is different from the temperature of the
medium:
 \begin{equation}
    f(p,t=0) = N\,e^{-p^2/2m T_0}.
 \end{equation}
In a medium at fixed temperature $T=100$~MeV and normal nuclear
density $\rho_0$ the solution $f(p,t)$ of the Fokker-Planck equation
can be calculated analytically:
 \begin{equation}
  \label{fpt}
    f(p,t) \sim \exp
    \left[ - \frac{e^{2 A t}}
                   {2D/A\left(e^{2 A t} - 1\right) + 2 m T_0}
            \, p^2 \right]
    \equiv \exp{\left[ - p^2/2 m T_{\rm eff}(t)\right] } \,.
 \end{equation}
This shows that the exponential shape of the distribution function is
maintained throughout the time evolution, but that the slope $T_{\rm
eff}(t)$ gradually evolves from $T_0$ to the value $D/mA$ which,
according to the Einstein relation (\ref{Einstein}), is the medium
temperature $T$. Looking at Fig.~\ref{Abb33} it is clear that after
about 10~fm/c the spectrum is practically thermalized.  Therefore
initial structures of the production spectrum (like the ones seen in
Fig.~\ref{Spektrum}) are washed out quite rapidely, and their
experimental observation will be very difficult.

%
%

\subsection{Annihilation}
\label{sec3B}

The annihilation of antiprotons with baryons is dominated by
multi-meson final states $X$. For the parametrisation of the
annihilation cross section $\sigma^{\rm ann}(s)$ we take the form
given in \cite{KD} for the process
 \begin{equation}
    \bar p + B \longrightarrow X\,,
    \qquad \mbox{ } \quad B=N,\,\Delta,\dots
 \label{barpB}
 \end{equation}

Using the same philosophy as for the calculation of the production
rate, a simple differential equation for the decrease of the
antiproton density in phase space can be written down:
 \begin{equation}
   \frac{d}{dt} \, \frac{d^6 N}{d^3 x d^3 p} = - \frac{d^6  N}{d^3 x
    \, d^3 p} \sum\limits_{i} \int \, \frac{d^3 p_i}{(2\pi)^3} \, f_i
    (\vec x,\,\vec p_i,\, t) \, v_{i\bar p} \, \sigma^{\rm ann}_{i\bar
    p} \equiv - \frac{d^6 N}{d^3 x \, d^3 p} \ A (\vec x, \vec p, t)
    \,.
 \label{ddt}
 \end{equation}
This equation shows that the absorption rate is given by the thermal
average of the annihilation cross section $A[\mu_B, T] = \langle v
\sigma^{\rm ann} \rangle$ where $v$ is the $\bar p$ velocity. Although
$\sigma^{\rm ann} (s)$ is strongly peaked for low-energy antiprotons,
the thermal average $\langle v \sigma^{\rm ann} \rangle$ is
practically independent of the $\bar p$ velocity in the medium.
Therefore, in a stationary medium the shape of the $\bar p$ spectrum
is maintained; the momentum independence of $A$ leads only to a
renormalisation of the spectrum by annihlation. Any modifications of
the spectral shape must be due to the dynamical evolution of the
hadronic medium.

The absolute value of $A$ decreases slightly with increasing
temperature; it is of the order of 0.5 (fm/$c$)$^{-1}$ at normal
nuclear matter density. This means that 99\% of the antiprotons are
absorbed within 9~fm/c. Of course, due to the rather large value of
the absorption rate $A$ which is of the order of the size of the
antiproton, the classical approach chosen in our model becomes
questionable. More reliable results should be based on a quantum field
theoretic calculation which is beyond the scope of this paper.

\section{Antiproton Spectra from an Exploding Fireball}
\label{sec4}

\subsection{A Model for the Heavy Ion Collision}
\label{sec4A}

In order to compare the results of the two previous sections with
experimental data we connect them through a dynamical model for the
heavy ion reaction. In the spirit of our thermodynamic approach the
so-called hadrochemical model of Montvay and Zimanyi \cite{MZ} is applied
for the simulation of the heavy ion collision. In this picture the
reaction is split into two phases, an ignition and an explosion phase,
and particles which have at least scattered once are assumed to follow
a local Maxwell-Boltzmann distribution. In addition, a spherically
symmetric geometry is assumed for the explosion phase. The included
particle species are nucleons, $\Delta$-resonances, pions and
$\rho$-mesons.

As initial condition a Fermi-type density distribution is taken for
the nucleons of the incoming nuclei,
 \begin{equation}
    \rho(r) = \frac{\rho_0}{1 + e^{(r - R_0)/z}}  \, ,
 \label{Woods}
 \end{equation}
with the nuclear radius $R_0$ and a surface diffuseness $z = 0.55$~fm.
Because the Fermi motion is small compared to the typical bombarding
energy of 1~GeV/A, it can be neglected, and the initial phase space
distribution therefore reads
 \begin{equation}
    f \arg = \delta(p_x)\,\delta(p_y)\,\delta(p_z-p_0)\
    \rho(|\vec x-\vec\xi(t)|)\,.
 \label{finit}
 \end{equation}
Here $\vec \xi (t)$ describes the path of the center of the nucleus.

\begin{figure}
  \epsfxsize=14cm
  \epsfbox{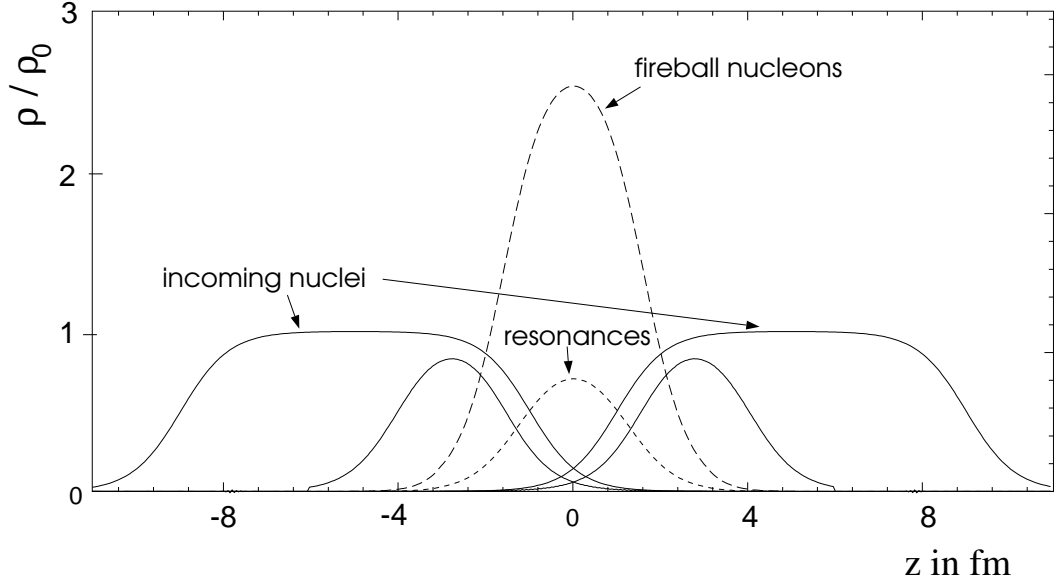}
  \vskip 0.4cm
  \caption{
    Density distributions $\rho(0, 0, z)$ along the beam axis of
    target and projectile nucleons for a $^{40}$Ca-$^{40}$Ca
    collision, normalized to $\rho_0 = 0.15$ fm$^{-3}$. The solid
    lines labelled by ``incoming nuclei'' show the two nuclei centered
    at $\pm 5$~fm at time $t_0=0$. The two other solid lines
    denote the cold nuclear remnants at full overlap time $t_m$,
    centered at about $\pm 3$ fm. Also shown are fireball nucleons
    (long-dashed) and $\Delta$-resonances (short-dashed) at time
    $t_m$.}
  \label{Vert90}
\end{figure}

Due to elastic and inelastic scattering hot and dense matter is formed
in the central region. Its time-dependent chemical composition is
governed by the kinetic equations of the hadrochemical model given in
Ref.~\cite{MZ}. Their particular form needed here can be found in
\cite{Diss}. Some of the resulting distributions are visualized in
Fig.~\ref{Vert90} for a central $^{40}$Ca-$^{40}$Ca collision with
momentum $p_z = \pm 700$~MeV in the c.m. system. One clearly sees that
at the moment of full overlap of the two nuclei a dense zone with
hot nucleons, resonances and mesons (not shown) has been formed. Only
in the peripheral regions "cold" target and projectile nucleons can
still be found. On the other hand, chemical equilibrium of the hot
collision zone is not reached in the short available time before the
explosion phase sets in; pions and in particular $\rho$ mesons
remain far below their equilibrium abundances \cite{Diss}.

%
%

It is important to note that due to the arguments given in
Section~\ref{sec2} $\bar p$ production is strongly suppressed in this
initial stage of the reaction. In our simple model we have in fact
neglected this early $\bar p$ production completely. The ignition
phase is only needed to obtain the chemical composition of the hot
fireball which is expected to be the dominant source for the creation
of antiprotons.

For the subsequent expansion of the spherically symmetric fireball
into the surrounding vacuum analytical solutions can be given if the
equation of state of an ideal gas is taken as input \cite{BGZ}. If
excited states are included in the model, an exact analytical solution
is no longer possible. Because a small admixture of resonaces is not
expected to fundamentally change the dynamics of the system, we can
account for their effect in first order by adjusting only the
thermodynamic parameters of the exploding fireball, but not the
expansion velocity profiles.

\begin{figure}
 \begin{center}
  \parbox{7cm}{$$\alpha = 0.2$$
    \epsfxsize=6.5cm\epsfbox{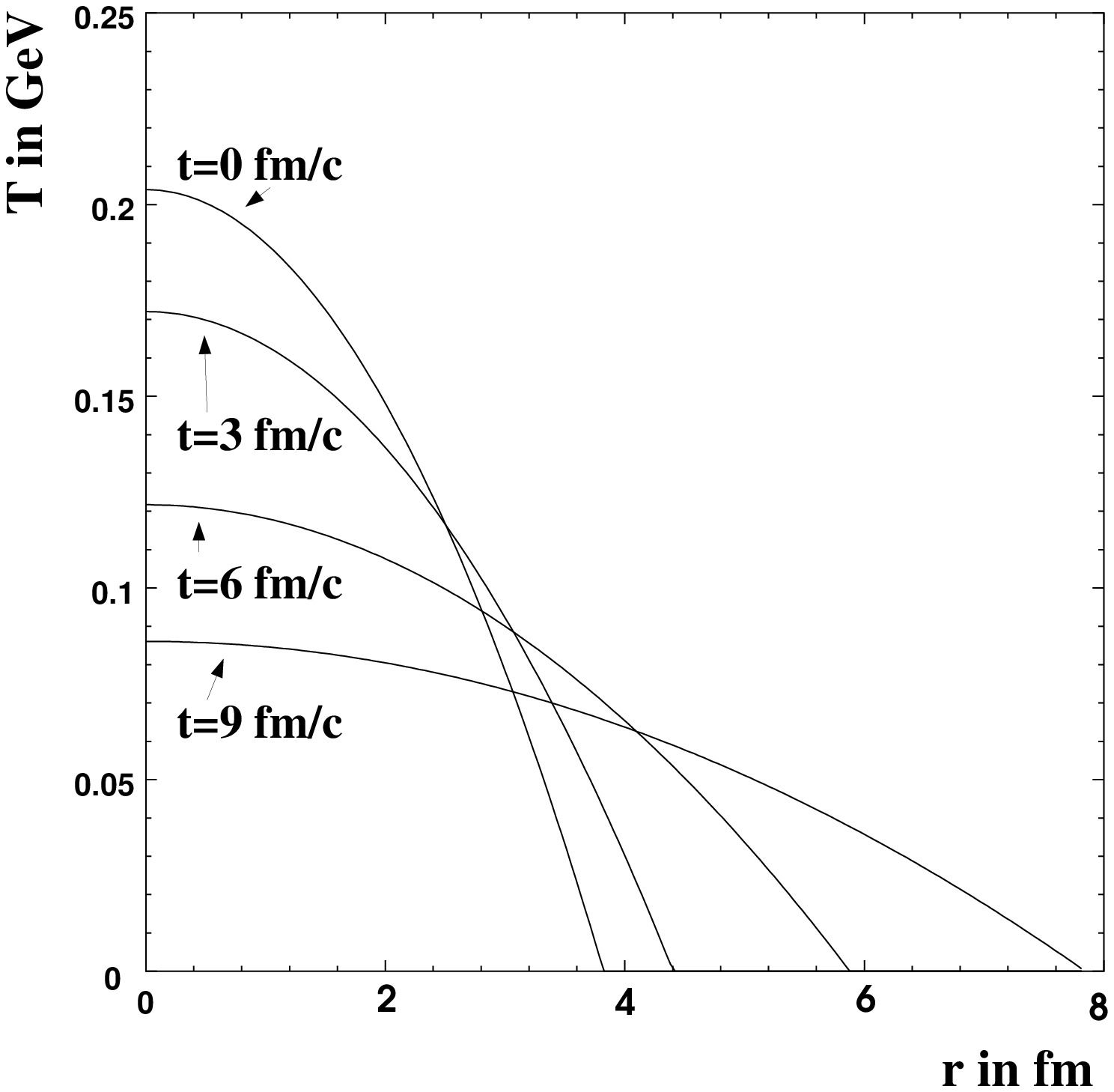}}
  \parbox{7cm}{$$\alpha = 5$$
    \epsfxsize=7cm\epsfbox{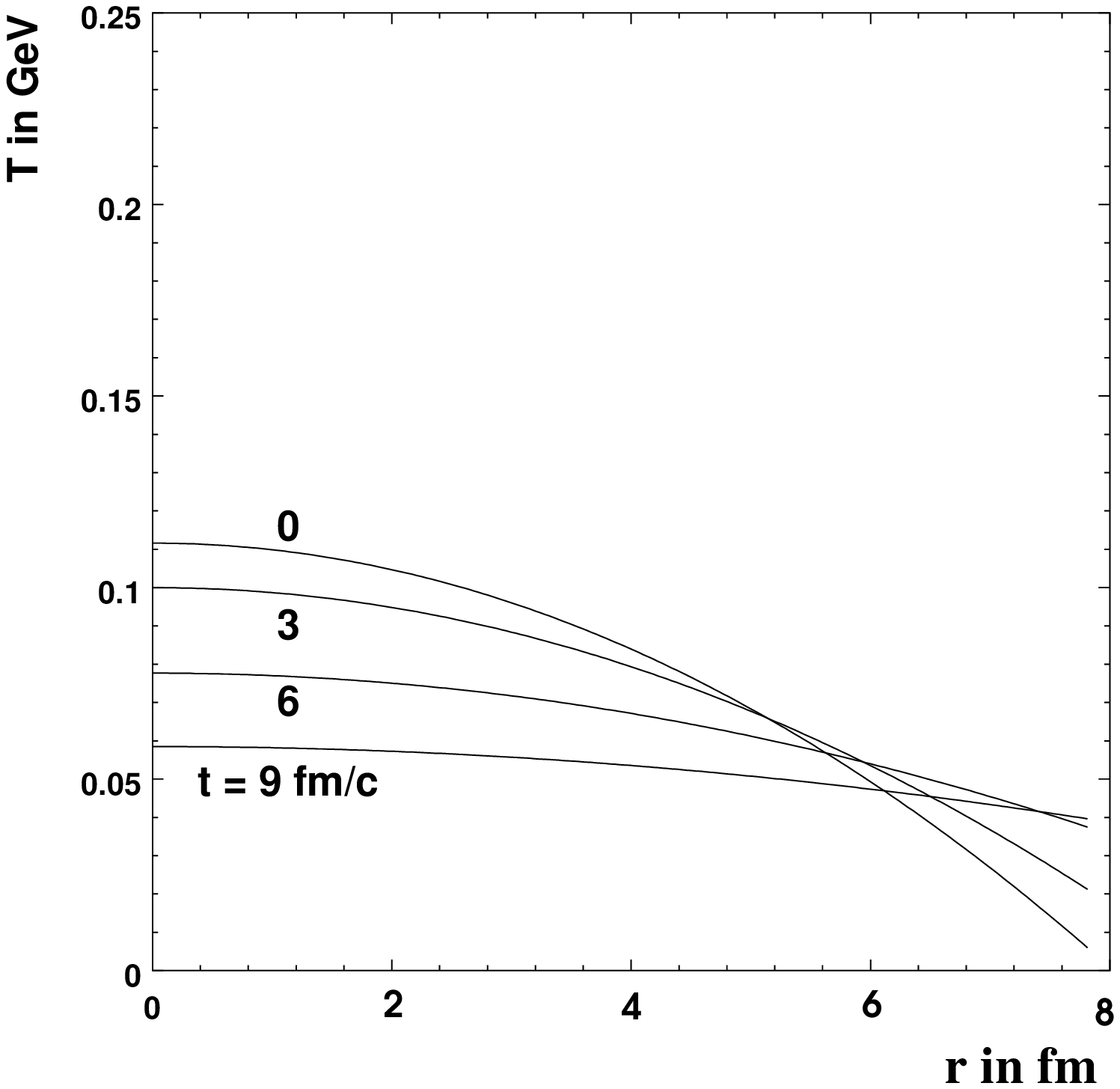}}
 \end{center}
 \vskip 0.4cm
 \caption{
    Temperature profiles for $\alpha=0.2$ and $\alpha= 5$ at four
    different times $t=0$ (=beginning of the explosion phase) and
    $t = 3,\  6,$ and 9~fm/$c$ (from top to bottom).}
 \label{Abb47}
\end{figure}

There is one free parameter in the model \cite{MZ}, $\alpha$, which
controls the density and the temperature profiles, respectively. Small
values $\alpha \rightarrow 0$ represent $\delta$-function like density
profiles whereas $\alpha \rightarrow \infty$ corresponds to a
homogeneous density distribution throughout the fireball (square well
profile). The time-dependent temperature profiles for two
representative values of $\alpha$ are shown in Fig.~\ref{Abb47} for
different times $t$ starting at the time $t_m$ of full overlap of the
nuclei. Clearly, a small value of $\alpha$ leads to an unreasonably
high temperature ($T \sim 200$~MeV) in the core of the fireball at the
beginning of the explosion phase, and should thus be considered
unphysical.

%
%

\subsection{Antiproton Spectra from an Exploding Fireball}
\label{sec4B}

\begin{figure}
  \epsfxsize8cm
  \epsfbox{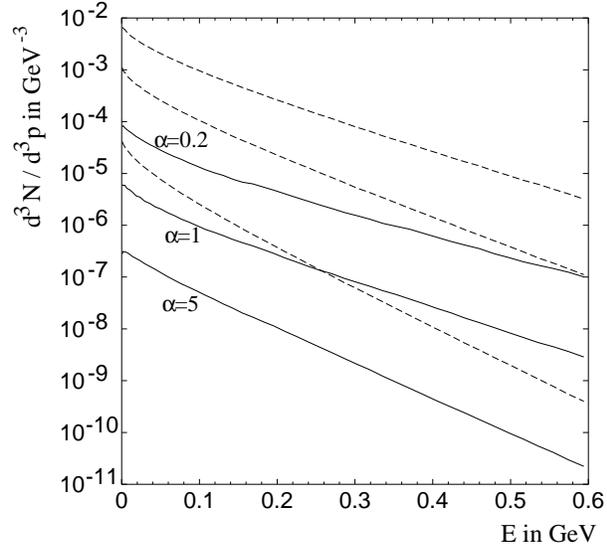}
  \vskip 0.4cm
  \caption{
     $\bar p$-spectra for different profile parameters $\alpha$. The
     dotted lines mark the initial production spectra. The asymptotic
     temperatures at an assumed freeze-out density $\rho_f = \rho_0
     /2$ corresponding to the solid lines are, from top to bottom,
     105~MeV, 87~MeV and 64~MeV, respectively.}
  \label{Abb54}
\end{figure}

Based on the time-dependent chemical composition of this hadrochemical
model we can calculate the spectrum of the antiprotons created
in a heavy ion collision. Let us first concentrate on the influence
of the density distribution in the fireball characterized by the shape
parameter $\alpha$. Due to different temperature profiles connected
with different $\alpha$ values (see Fig.~\ref{Abb47}) the absolute
normalization varies substantially when the density profile is
changed. For more $\delta$-like shapes an extremely hot fireball core
is generated, whereas for increasing $\alpha$ both density and
temperature are more and more diffuse and spread uniformly over a
wider area. Because of the exponential dependence upon temperature a
small, but hot core raises the production rate drastically. This fact
is illustrated in Fig.~\ref{Abb54} for three values of $\alpha$. Not
only the total normalization, but also the asymptotic slope of the
spectrum is modified due to the variation of the core temperature with
$\alpha$ as indicated in the Figure.

%
%

Comparing the dotted lines, which give the pure production spectrum, to
the solid lines representing the asymptotic spectrum at decoupling,
the tremendous effect of antiproton absorption in heavy ion collisions
is obvious. As one intuitively expects absorption is more pronounced
for low-energetic antiprotons than for the high-energetic ones which
have the opportunity to escape the high density zone earlier.
Therefore, the finally observed spectrum is flatter than the original
production spectrum.

Interestingly, while the baryon-baryon and the $\rho\rho$ channels are
comparable in their contribution to $\bar p$ production, the pion-baryon
channel turned out to be much more effective for all reasonable sets of
parameters. This fact is indeed remarkable, because here, contrary to
the discussion in Section~\ref{sec2}, the pions are not in chemical
equilibrium; in our hadrochemical model the total time of the ignition
phase is too short to saturate the pion channel. The meson-baryon
channel is thus crucial for understanding $\bar p$ spectra. Only by
including all channels reliable predictions about the antiprotons can
be drawn.

\begin{figure}
  \epsfysize7.5cm
  \epsfbox{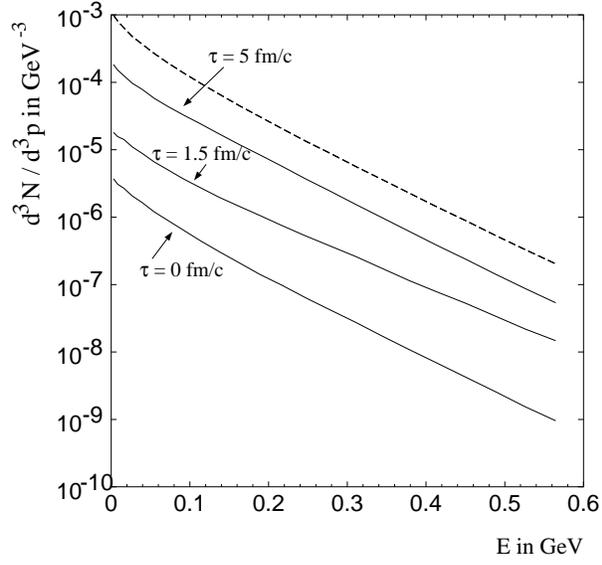}
  \vskip 0.4cm
  \caption{
      $\bar p$-spectrum for different formation times, for a profile
      parameter $\alpha = 1$. The dashed line indicates the original
      production spectrum.}
  \label{Abb55}
\end{figure}

We did not mention so far that in our calculations we followed common
practice and assumed a finite $\bar p$ formation time of $\tau =
1$~fm/c; this means that during this time interval after a
$\bar p$-producing collision the antiproton is assumed to be not yet
fully developed and thus cannot annihilate. However, there are some
(although controversial) experimental indications of an extremely long
mean free path for antiprotons before final state interactions set in
\cite{Va}. We have tested the influence of different values for the
formation time $\tau$ on the $\bar p$ spectrum. Fig.~\ref{Abb55} shows
that this highly phenomenological and poorly established parameter has
a very strong influence in particular on the absolute normalization of
the spectra, i.e. the total production yield. In the light of this
uncertainty it appears difficult to argue for or against the necessity
for medium effects on the antiproton production threshold based on a
comparison between theoretical and experimental total yields only.

%
%

\subsection{Comparison with Experimental Data}
\label{sec4C}

In  all the calculations shown above a bombarding energy of 1~GeV/A
has been assumed. Experimental data are, however, only available at
around 2~GeV/A. At these higher energies thermalization becomes more
questionable \cite{La}, and our simple model may be stretching its
limits. Especially, the temperature in the fireball core becomes
extremely high. In order to avoid such an unrealistic situation and in
recognition of results from kinetic simulations \cite{TCMM,Bat,Sp} we
thus assume that only part of the incoming energy is thermalized -- in
the following a fraction of 70\% was taken.

%
%

\begin{figure}
   \epsfxsize8cm
   \epsfbox{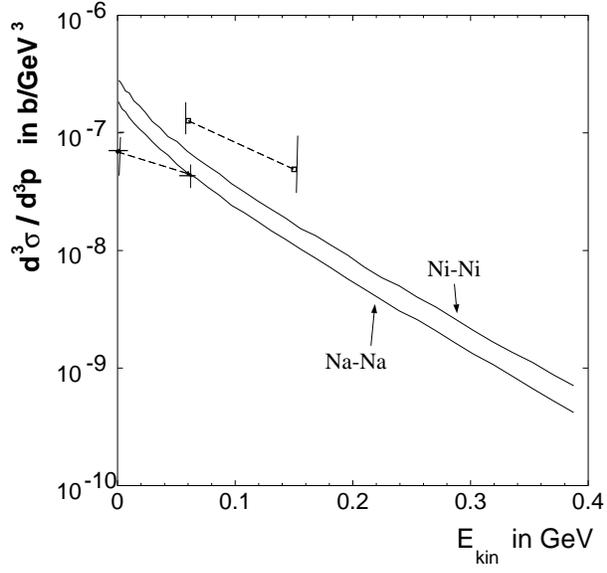}
   \vskip 0.4cm
   \caption{
     Differential $\bar p$ spectrum for Na-Na and Ni-Ni collisions,
     for a shape parameter $\alpha=1$. The data are from GSI
     experiments \protect\cite{Schr}.}
   \label{Bild111}
\end{figure}

Fig.~\ref{Bild111} shows calculations for the antiproton spectrum from
Na-Na and Ni-Ni collisions at a kinetic beam energy of 2 GeV/nucleon.
The calculation assumes a $\bar p$ formation time of $\tau=1$ fm/$c$,
and takes for the density and temperature profiles the parameter value
$\alpha=1$ which corresponds to an upside-down parabola for the
density profile. Comparing with the GSI data \cite{Schr} we see our
model features too weak a dependence on the size of the collision
system; the absolute order of magnitude of the antiproton spectrum is,
however, correctly reproduced by our simple hadrochemical model,
without adjusting any other parameters. No exotic processes for $\bar
p$ production are assumed. As mentioned in the previous subsection,
the pion-baryon channel is responsible for getting enough antiprotons
in our model, without any need for a reduced effective $\bar p$ mass
in the hot and dense medium \cite{TCMM,LK}. The existing data do not
yet allow for a definite conclusion about the shape of the spectrum,
and we hope that future experiments \cite{Gill} will provide
additional contraints for the model.

\section{Conclusions}
\label{sec5}

\begin{figure}
 \epsfxsize=10cm
 \epsfbox{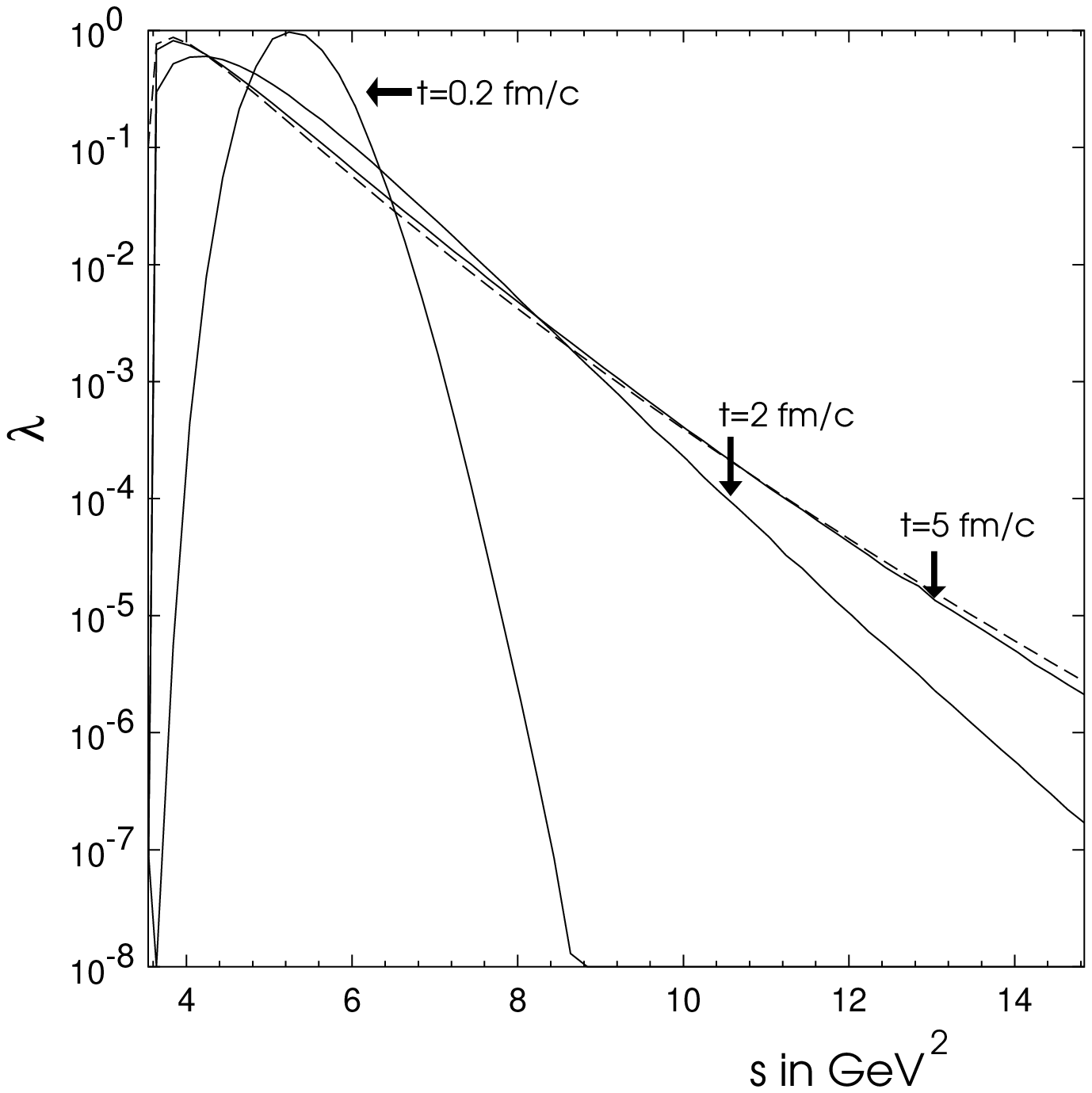}
 \vskip 0.4cm
 \caption{
   $\lambda(s,t)$ at different times $t$ calculated from the
   model of Ref.~\protect\cite{SHP}. The starting point is a
   $\delta$-function at $s=5.5\,\mbox{GeV}^2$. The dashed line is the
   asymptotic thermal distribution at $t{=}\infty)$, corresponding to
   a temperature $T=133$~MeV.}
\label{Bild1}
\end{figure}

Heavy ion collisions at typical BEVALAC and SIS energies are far below
the $p\bar p$-production threshold. As a consequence, pre-equilibrium
antiproton production in such collisions is strongly suppressed
relative to production from the thermalized medium produced in the
later stages of the collision. Therefore, $\bar p$ production becomes
important only when the heavy ion reaction is sufficiently far
progressed, in accordance with microscopic simulations \cite{TCMM}. By
assuming a local Maxwell-Boltzmann distribution for the scattered and
produced particles forming the medium in the collision zone one
maximizes the $\bar p$ production rate (see Fig.~\ref{Bild1}). If,
contrary to the assumptions made in this work, the extreme states in
phase space described by the tails of the thermal Boltzmann
distribution are not populated, the antiproton yield could be reduced
substantially.

We also found that the threshold behaviour of the $\bar p$ production
cross section is not only crucial for the total $\bar p$ yield, but
also introduces structures into the initial $\bar p$ spectrum. This
might give rise to the hope that by measuring the $\bar p$ momentum
spectrum one may obtain further insight into the $\bar p$ production
mechanism. On the other hand we saw here, using a Fokker-Plank
description for the later evolution of the distribution function
$f_{\bar p}(p,t)$ in a hot environment, that these structures are
largely washed out by subsequent elastic scattering of the $\bar p$
with the hadrons in the medium. In addition, the large annihilation
rate reduces the number of observable antiprotons by roughly two
orders of magnitude relative to the initial production spectrum; the
exact magnitude of the absorption effect was found to depend
sensitively on the choice of the $\bar p$ formation time $\tau$.

We have also shown that meson (in particular pion) induced production
channels contribute significantly to the final $\bar p$ yield and
should thus not be neglected. We were thus able to reproduce the total
yield of the measured antiprotons in a simple model for the reaction
dynamics without including, for example, medium effects on the hadron
masses and cross sections \cite{TCMM,LK}.

However, we must stress the strong sensitivity of the $\bar p$ yield
on various unknown parameters (e.g. the $\bar p$ formation time) and
on poorly controlled approximations (e.g. the degree of population of
extreme corners in phase space by the particles in the collision
region), and emphasize the rapidly thermalizing effects of elastic
final state interactions on the $\bar p$ momentum spectrum. We
conclude that turning subthreshold antiproton production in heavy ion
collisions into a quantitative probe for medium properties and
collective dynamics in hot and dense nuclear matter remains a serious
challenge.

\acknowledgements

This work was supported by the Gesellschaft f\"ur Schwerionenforschung
(GSI) and by the Bundesministerium f\"ur Bildung und Forschung (BMBF).



\newpage

\newpage

\newpage

\newpage

\newpage

\newpage

\newpage

\end{document}